\documentclass[journal]{IEEEtran}

\ifCLASSINFOpdf
\else
   \usepackage[dvips]{graphicx}
\fi
\usepackage{url}

\usepackage{graphicx}

\usepackage{amsmath}
\usepackage{amssymb}
\usepackage{pifont}
\usepackage[ruled,vlined]{algorithm2e}
\usepackage{array}
\usepackage{multirow}
\usepackage[table]{xcolor}
\usepackage{microtype}
\usepackage{etoolbox}
\usepackage{xcolor}

\newcommand{\mcolor}{black}   
\newcommand{\rev}[1]{\textcolor{\mcolor}{#1}}
\newcommand{\revaq}[1]{\textcolor{black}{#1}}

\AtEndEnvironment{algorithm}{\vspace{-0pt}}

\begin{document}

\title{Sound Event Detection with \\ Boundary-Aware Optimization and Inference}

\author{Florian Schmid, Chi Ian Tang, Sanjeel Parekh, Vamsi Krishna Ithapu, Juan Azcarreta Ortiz, Giacomo Ferroni, Yijun Qian, Arnoldas Jasonas, Cosmin Frateanu, Camilla Clark, Gerhard Widmer, Çağdaş Bilen%
\thanks{F.~Schmid conducted this work during an internship at Meta. He and G.~Widmer are affiliated with the Institute of Computational Perception, with G.~Widmer also at the Linz Institute of Technology (LIT). The other authors are with Meta Reality Labs Research.
}
}

\maketitle


\begin{abstract}

Temporal detection problems appear in many fields including time-series estimation, activity recognition and sound event detection (SED). In this work, we propose a new approach to temporal event modeling by explicitly modeling event onsets and offsets, and by introducing boundary-aware optimization and inference strategies that substantially enhance temporal event detection. The presented methodology incorporates new temporal modeling layers—Recurrent Event Detection (RED) and Event Proposal Network (EPN)—which, together with tailored loss functions, enable more effective and precise temporal event detection. We evaluate the proposed method in the SED domain using a subset of the temporally-strongly annotated portion of AudioSet. Experimental results show that our approach not only outperforms traditional frame-wise SED models with state-of-the-art post-processing, but also removes the need for post-processing hyperparameter tuning, and scales to achieve new state-of-the-art performance across all AudioSet Strong classes.

\end{abstract}


\IEEEpeerreviewmaketitle

\vspace{-5pt}

\section{Introduction}

Automatically identifying and interpreting sounds in real-world environments is essential for applications ranging from smart homes~\cite{debes2016monitoring} and healthcare monitoring~\cite{zigel2009method} to security and surveillance~\cite{radhakrishnan2005audio}. Audio recognition tasks extract information at different granularities: audio tagging~\cite{virtanen2018computational} catalogs events at clip level, while sound event detection (SED)~\cite{virtanen2018computational,mesaros2021sound} further identifies event types and their precise temporal boundaries. This enables detailed reconstruction of event sequences, durations, and overlaps, providing a richer understanding of complex acoustic scenes and allowing for downstream tasks, such as event-based audio editing~\cite{Singh04Event-basedAE}. 

Formally, SED aims to detect a set of events $E = \{e_j\}$, where each event $e_j = (c_j, t^{\mathrm{start}}_j, t^{\mathrm{end}}_j)$ consists of the event class $c_j$, start time $t^{\mathrm{start}}_j$, and end time $t^{\mathrm{end}}_j$. This work focuses on improving the accuracy of event boundary detection, i.e., the precise estimation of $t^{\mathrm{start}}_j$ and $t^{\mathrm{end}}_j$. While a few SED systems have been developed to directly predict a set of events $\hat{E}$~\cite{Venkatesh22Yoho,ye2021sound, bhosale2024diffsed}, most models—including current state-of-the-art approaches~\cite{schmid2025effective,cai2024mat,nam2022frequency,Schmid2024multised}—output frame-level scores that require post-processing to obtain the final event set. This dominance is largely due to the optimization benefits of frame-wise models: they naturally support multiple instance learning~\cite{kumar2016audio,shah2018closer,wang2019comparison}, enabling straightforward training on weakly labeled data (without precise temporal annotations), and allow for simple, low-complexity architectures such as the commonly used \revaq{convolutional recurrent neural network (CRNN)}~\cite{cakir2017convolutional,xu2018large,li2020sound}.

Although frame-wise models are widely used, they have notable limitations. Typically trained with frame-wise binary cross-entropy \revaq{(BCE)} loss, these models focus on frame-level accuracy but fail to capture event continuity. Events are constructed by thresholding frame-level scores and grouping consecutive frames above the threshold. To reduce temporal fluctuations, post-processing—most commonly median filtering (MF)~\cite{cai2024mat,shao2024fine,nam2022frequency,schmid2025effective,turpault2019sound}—is applied. Recently, Sound Event Bounding Boxes (SEBB)~\cite{ebbers2024sebbs} has emerged as a state-of-the-art post-processing method, substantially outperforming MF by decoupling event region detection from thresholding. However, both MF and SEBB are non-differentiable and require hyperparameter tuning on a validation set, separate from model training, which can lead to suboptimal performance~\cite{yoshinaga2024onset}. \revaq{A closely related approach to ours is \textit{HSM3}~\cite{yoshinaga2024onset}, a hidden semi-Markov model-based method that also performs end-to-end event inference on frame-wise models. However, HSM3 relies on explicit duration modeling and forward--backward inference, whereas our method directly predicts event regions, yielding a more lightweight formulation.} The proposed method in this paper combines the strengths of both major SED paradigms, end-to-end event prediction and frame-wise modeling:


\textbf{Direct Event Region Prediction:} Inspired by Region Proposal Networks~\cite{girshick2014rpn,girshick2015fastrpn,ren2015fasterrpn} in computer vision—which predict spatial object locations—we introduce Event Proposal Networks (EPNs) that directly predict temporal event locations.

\textbf{No Post-processing Hyperparameters:} Unlike MF or SEBB, our approach removes the requirement for tuning post-processing hyperparameters after training.


\textbf{Model Flexibility:} Our method extends frame-wise models while preserving architectural flexibility (from CRNNs to Transformers), and does not require encoder-decoder designs or matching algorithms (e.g., Hungarian matching~\cite{kuhn1955hungarian}) during training, unlike other end-to-end SED approaches~\cite{ye2021sound,bhosale2024diffsed}.

\rev{As frame-wise SED models currently achieve top performance on major benchmarks (DESED~\cite{turpault2019sound}, AudioSet Strong~\cite{hershey2021benefit}), our evaluation focuses on frame-wise architectures. We evaluate our approach on ten short-duration classes from AudioSet Strong (AS-Strong)~\cite{hershey2021benefit}, where accurate boundary detection is critical.}
Our method yields substantial improvements over traditional frame-wise models with SEBB or \textit{HSM3} post-processing. When scaling to all AS-Strong classes, our approach achieves a new state-of-the-art PSDS1 score~\cite{bilen2020framework,ebbers2022threshold} of 49.6, surpassing the previous best of 46.5~\cite{schmid2025effective}.



\vspace{-5pt}
\section{Method}
\label{sec:method}

We begin by providing an overview of our proposed method, followed by detailed descriptions of each component. Fig.~\ref{fig:overview} illustrates the overall system, its outputs, and their connections to the various loss functions. The \revaq{recurrent event detection (RED) layer} (Section~\ref{subsec:red_layer}), placed atop any frame-wise acoustic model, converts conditional event start and end probabilities into onset, offset, and event presence probabilities, enabling direct training on onsets and offsets (Section~\ref{subsec:ool}). Rather than relying on post-processing to convert presence probabilities into events, we introduce event proposal networks (Section~\ref{subsec:event_proposal_networks}), which use RED outputs to generate frame-wise duration estimates and establish event region proposals. Finally, for inference, we select the most suitable proposals using a non-maximum suppression-like algorithm (Section~\ref{subsec:inference_algo}). 

\vspace{-10pt}
\subsection{RED Layer}
\label{subsec:red_layer}

The recurrent event detection layer\footnote{RED was first introduced by the authors in~\cite{Bilen2023patent}. The present paper provides the first formal and detailed academic introduction of RED.} models event onset, offset, and presence probabilities via a \rev{parameter-less, differentiable} probabilistic recurrent relationship. RED can be added to any acoustic model with frame-wise outputs, requiring only a minor change: instead of a single output per class (event presence), the model outputs two values per class—one for event start and one for event end at each frame. The RED formulation uses a single random variable $E_{c,t}$, representing event presence at frame $t$ for class $c$. Since RED operates independently for each class, we omit the class index for clarity. The inputs to RED are the estimated conditional event start $P(e_t \mid \neg e_{t-1})$ and event end $P(\neg e_t \mid e_{t-1})$ probabilities, obtained by applying a sigmoid to the frame-wise acoustic model output logits. RED computes frame-wise event presence probability using the following probabilistic recurrence:


\vspace{-10pt}
{\small
\begin{equation}
\underbrace{P(e_t)}_{\text{Pres. Prob.}} =
\underbrace{P(e_t \mid \neg e_{t-1})}_{\text{Event Start Prob.}} \cdot
\underbrace{P(\neg e_{t-1})}_{\text{Prev. Frame}}
+\bigl[1-\underbrace{P(\neg e_t \mid e_{t-1})}_{\text{Event End Prob.}}\bigr] \cdot
\underbrace{P(e_{t-1})}_{\text{Prev. Frame}}
\label{eq:red_recurrence}
\end{equation}
}

This formulation enables direct computation of onset and offset probabilities:

{\small
\begin{equation}
\begin{aligned}
\underbrace{P(e_t, \neg e_{t-1})}_{\text{Onset Prob.}}
    &= \underbrace{P(e_t \mid \neg e_{t-1})}_{\text{Event Start Prob.}}
       \cdot \underbrace{P(\neg e_{t-1})}_{\text{Prev. Frame}} \\[1em]
\underbrace{P(\neg e_t, e_{t-1})}_{\text{Offset Prob.}}
    &= \underbrace{P(\neg e_t \mid e_{t-1})}_{\text{Event End Prob.}}
       \cdot \underbrace{P(e_{t-1})}_{\text{Prev. Frame}}
\end{aligned}
\label{eq:onset_offset}
\end{equation}
}

\begin{figure}
\centerline{\includegraphics[width=\columnwidth]{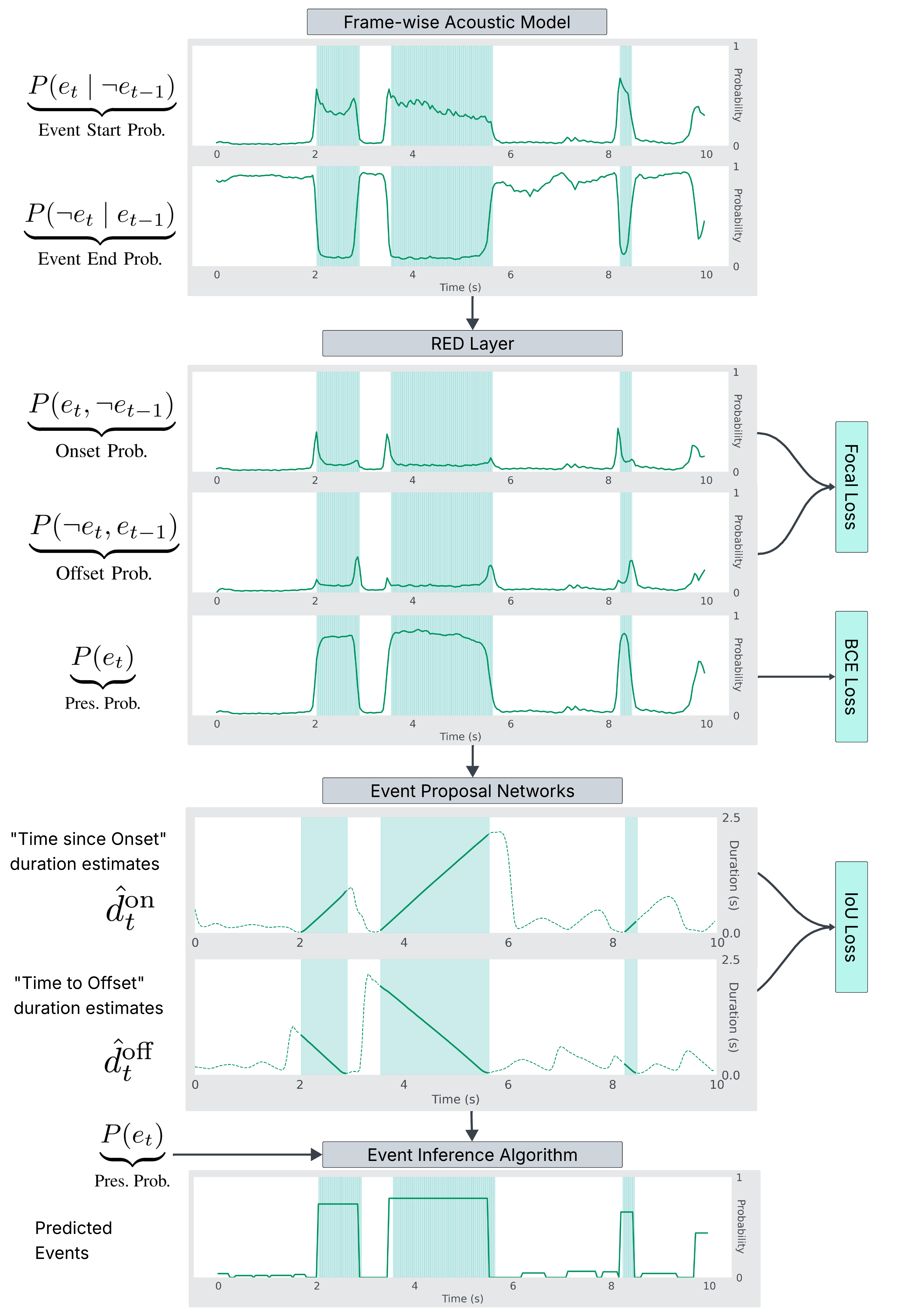}}
\caption{Example for the class \textit{Vehicle Horn} with three active ground truth events (colored boxes). Predicted probabilities and duration estimates are shown as line plots, linked to their respective loss functions.}
\label{fig:overview}
\end{figure}

Fig.~\ref{fig:overview} illustrates the distinction between conditional event start/end probabilities and onset/offset probabilities: while the conditionals act as simple "switch on/off" signals, the onset and offset probabilities are temporally localized peaks indicating event boundaries. We denote the per-class frame-wise presence, onset, and offset probabilities as $\hat{p}^{\mathrm{pres}}_{c,t}$, $\hat{p}^{\mathrm{on}}_{c,t}$, and $\hat{p}^{\mathrm{off}}_{c,t}$, respectively. RED can be efficiently parallelized using Heinsen scan~\cite{heinsen2023scan}, making its computational overhead negligible.

\vspace{-5pt}
\subsection{Onset-Offset-Loss \revaq{(OOL)}}
\label{subsec:ool}

\rev{Given the onset and offset probabilities exposed by RED, we train the underlying acoustic frame-wise model directly on ground-truth event boundaries.} From the event annotations, we derive frame-wise onset and offset labels $y^{\mathrm{on}}_{t,c}$ and $y^{\mathrm{off}}_{t,c}$, which are one at frames where an onset or offset occurs, and zero otherwise. Due to the sparsity and impulse-like nature of these labels, focal loss~\cite{Lin2017focal} is particularly effective. The resulting loss, with $\alpha=2$ in our setup, is:

\begin{equation}
\small
\begin{aligned}
\mathcal{L}^{\mathrm{on,off}} = -\frac{1}{CT} \sum_{c=1}^{C} \sum_{t=1}^{T}
\begin{cases}
    (1-\hat{p}^{\mathrm{on,off}}_{c,t})^{\alpha} \log \hat{p}^{\mathrm{on,off}}_{c,t} & \text{if } y^{\mathrm{on,off}}_{c,t} = 1 \\
    (\hat{p}^{\mathrm{on,off}}_{c,t})^{\alpha} \log (1-\hat{p}^{\mathrm{on,off}}_{c,t}) & \text{if } y^{\mathrm{on,off}}_{c,t} = 0
\end{cases}
\end{aligned}
\label{eq:onoff_loss}
\end{equation}

Since RED tightly couples $\hat{p}^{\mathrm{pres}}_{c,t}$, $\hat{p}^{\mathrm{on}}_{c,t}$, and $\hat{p}^{\mathrm{off}}_{c,t}$, applying losses to $\hat{p}^{\mathrm{on}}_{c,t}$ and $\hat{p}^{\mathrm{off}}_{c,t}$ directly influences the shape of $\hat{p}^{\mathrm{pres}}_{c,t}$.




\vspace{-5pt}
\subsection{Event Proposal Networks}
\label{subsec:event_proposal_networks}


While it is possible to apply standard post-processing to the refined $\hat{p}^{\mathrm{pres}}_{c,t}$ to extract events, this approach requires tuning post-processing hyperparameters on a separate validation set, decoupled from model training, which can result in suboptimal performance. To address this, we instead aim to learn event region proposals end-to-end during training, introducing \textit{Event Proposal Networks (EPNs)}. We employ two-layer bidirectional GRUs that operate directly on the frame-wise probabilities $\hat{p}^{\mathrm{pres}}_{c,t}$, $\hat{p}^{\mathrm{on}}_{c,t}$, and $\hat{p}^{\mathrm{off}}_{c,t}$ and consider two strategies:

\textbf{Per-Class GRUs:} For each class, we stack the probabilities along the channel dimension, \rev{resulting in $|C|$ class-specific GRUs}, each processing inputs of shape \rev{$[0,1]^{3 \times T}$}.

\textbf{Single GRU:} We stack all class-wise probabilities along the channel dimension, yielding a single GRU that processes inputs of shape \rev{$[0,1]^{3|C| \times T}$} ($3|C|$ channels, $T$ frames).

Each class-wise GRU produces outputs in \rev{$(0,\infty)^{T \times 2}$ (or $(0,\infty)^{T \times |C| \times 2}$} for the single GRU variant), yielding two duration estimates per time frame. These correspond to the \textit{time since event onset} ($\hat{d}^{\mathrm{on}}_{c,t}$) and the \textit{time to next event offset} ($\hat{d}^{\mathrm{off}}_{c,t}$). To ensure non-negative durations, we apply a Softplus activation.

During training, we optimize the duration estimates on all active frames (i.e., frames with ongoing events). For these frames, we extract ground truth durations $d^{\mathrm{on}}_{c,t}$ and $d^{\mathrm{off}}_{c,t}$, and construct the corresponding intervals $r_{c,t} = [t - d^{\mathrm{on}}_{c,t},\, t + d^{\mathrm{off}}_{c,t}]$ and $\hat{r}_{c,t} = [t - \hat{d}^{\mathrm{on}}_{c,t},\, t + \hat{d}^{\mathrm{off}}_{c,t}]$ for ground truths and predictions\footnote{For simplicity, we use $t$ to denote both the time and the frame index, related by the linear mapping $t = f \cdot \Delta t$, where $t$ is the time, $f$ is the frame index and $\Delta t$ is the frame duration.}, respectively. The IoU loss, using the binary event presence labels $y^{\mathrm{pres}}_{c,t}$, is defined as:

{\small
\begin{equation}
\mathcal{L}^{\mathrm{iou}} = 
    \frac{1}{\sum_{c} \sum_{t} y^{\mathrm{pres}}_{c,t}}
    \sum_{c} \sum_{t} 
    y^{\mathrm{pres}}_{c,t} \cdot 
    \frac{1 - \mathrm{IoU}(r_{c,t}, \hat{r}_{c,t})}
         {d^{\mathrm{on}}_{c,t} + d^{\mathrm{off}}_{c,t}}
\end{equation}
}

This loss is particularly effective, as $r_{c,t}$ and $\hat{r}_{c,t}$ always overlap, and it consistently outperformed direct regression on duration estimates. Weighting by event duration ensures equal loss contribution from all events, regardless of their length.

\revaq{The final loss formulation is a weighted sum of the losses introduced in this section and the standard frame-wise BCE loss $\mathcal{L}^{\mathrm{pres}}$ on the presence probabilities $\hat{p}^{\mathrm{pres}}_{c,t}$, resulting in a combined objective:}

\begin{equation}
\small
\mathcal{L}^{\mathrm{total}} = \mathcal{L}^{\mathrm{pres}} + \lambda_{\mathrm{ool}} \left( \mathcal{L}^{\mathrm{on}} + \mathcal{L}^{\mathrm{off}} \right) + \lambda_{\mathrm{iou}} \mathcal{L}^{\mathrm{iou}}
\end{equation}

Throughout all experiments, we set $\lambda_{\mathrm{ool}}=100$ and $\lambda_{\mathrm{iou}}$ is treated as a tunable hyperparameter.

\begin{algorithm}[t]
\small
\caption{Event Inference Algorithm}

\KwIn{Pres. Probs. $\hat{p}^{\mathrm{pres}}_{c,t}$, Reg. Prop. $\hat{r}_{c,t}$, $k$ (max events per class), $m$ (max classes)}

\KwOut{Events $\hat{\mathcal{E}} = \{(c_j, \hat{t}^{\mathrm{start}}_j, \hat{t}^{\mathrm{end}}_j, \hat{\sigma}_j)\}$}

Compute $\bar{p}^{\mathrm{pres}}_c = \mathrm{mean}_t(\hat{p}^{\mathrm{pres}}_{c,t})$ for all $c$\;

Select $m$ classes with highest $\bar{p}^{\mathrm{pres}}_c$\;

\ForEach{selected class $c$}{
    Sort $\hat{r}_{c,t}$ by $\hat{p}^{\mathrm{pres}}_{c,t}$ in descending order\;
    Initialize $\hat{\mathcal{E}}_c \leftarrow \emptyset$\;
    \While{proposals remain \textbf{and} $|\hat{\mathcal{E}}_c| < k$}{
        Select top proposal $\hat{r}_{c,t^*}$\;
        $\hat{\sigma} = \mathrm{mean}(\hat{p}^{\mathrm{pres}}_{c,t})$ over $t \in \hat{r}_{c,t^*}$\;
        Add $(c, t^* - \hat{d}^{\mathrm{on}}_{c,t^*}, t^* + \hat{d}^{\mathrm{off}}_{c,t^*}, \hat{\sigma})$ to $\hat{\mathcal{E}}_c$\;
        Remove proposals overlapping with $\hat{r}_{c,t^*}$\;
    }
}
\Return $\hat{\mathcal{E}} = \bigcup_c \hat{\mathcal{E}}_c$\;

\label{alg:region_prop_inference}
\end{algorithm} 

\vspace{-5pt}
\subsection{Event Inference Algorithm}
\label{subsec:inference_algo}

\rev{Alg.~\ref{alg:region_prop_inference} selects relevant frame-wise region proposals and converts them into event predictions via non-maximum suppression, using the event-presence probabilities $\hat{p}^{\mathrm{pres}}_{c,t}$ from RED and the region proposals $\hat{r}_{c,t}$ from the EPNs.} For efficiency, we introduce two parameters: $k$, the maximum number of expected events per recording, and $m$, the number of most active classes (based on $\hat{p}^{\mathrm{pres}}_{c,t}$) considered during inference. Lower values of $k$ and $m$ reduce runtime at the potential cost of performance. By default, we set $k=15$ and $m=|C|$. Fig.~\ref{fig:overview} visualizes the inputs ($\hat{p}^{\mathrm{pres}}_{c,t}$, $\hat{r}_{c,t}$) and the corresponding events generated by Alg.~\ref{alg:region_prop_inference}.

\vspace{-5pt}
\section{Experimental Setup}

\subsection{Dataset \& Metrics}
\label{subsec:dataset}

We conduct experiments on AS-Strong~\cite{hershey2021benefit}, the largest publicly available dataset with strong temporal audio annotations. Following~\cite{schmid2025effective}, our training and evaluation sets comprise 100{,}911 and 16{,}935 10-second audio clips, respectively. AS-Strong contains 447 classes, with 407 present in both training and evaluation. The dataset is highly imbalanced, with many rare classes represented by only a few event instances. AS-Strong does not provide a predefined validation split, and the abundance of rare classes makes it difficult to construct a representative validation set.


\revaq{To facilitate analysis, we first focus on 10 well-defined classes selected for sufficient train/evaluation representation and short average event durations (at most 3 seconds), making precise temporal localization particularly important.} The chosen classes are \textit{Alarm}, \textit{Bark}, \textit{Cough}, \textit{Explosion}, \textit{Gunshot}, \textit{Laughter}, \textit{Screaming}, \textit{Vehicle horn}, \textit{Whispering}, and \textit{Whistling}. The resulting \textit{AS-Strong-10} subset contains 15,829 training and 2,394 evaluation files. This setup enables a well-defined validation set via a multilabel stratified 80:20 train/validation split, before scaling to the full AudioSet (\textit{AS-Strong-Full}).

Our primary evaluation metric is the threshold-independent PSDS1 score~\cite{ebbers2022threshold} (\textit{P1}), the standard for temporally strict SED assessment~\cite{cornell2024dcase}. Following~\cite{li2024atst-f,schmid2025effective}, we omit the variance penalty in PSDS1 computation. As a complementary metric, we report the collar-based F1 score (\textit{F1}) with a 200\,ms tolerance, where the allowed offset deviation is $\max(\text{offset\_collar},\, 0.2 \times \text{event length})$, ensuring the tolerance scales with event duration.

\vspace{-5pt}
\subsection{Architectures \& Training}

\revaq{Our method is designed to operate on any frame-wise SED architecture, replacing temporal post-processing. We evaluate on the CRNN baseline~\cite{cakir2017convolutional,xu2018large,li2020sound}, the state-of-the-art transformer models ATST-F~\cite{li2024atst-f} and BEATs~\cite{chen2022beats}, and the efficient MobileNetV3+GRU (MN-GRU) model~\cite{morocutti2025exploring}. These models were selected to cover a broad range of commonly used SED architectures, from lightweight CRNN-based systems to large-scale transformer models.} ATST-F, BEATs, and MN-GRU are pre-trained on AudioSet weak labels~\cite{gemmeke2017audio}, while CRNN is trained from scratch. 
On \textit{AS-Strong-10}, all models are trained with a batch size of 128 for up to 100 epochs, using the AdamW optimizer~\cite{loshchilov2017adamw} (weight decay 1e-3) and a cosine learning rate schedule with 1,000 warmup steps. The maximum learning rate is tuned per model. Data augmentation includes Freq-MixStyle~\cite{kim2022domain,schmid2022knowledge}, filter augmentation~\cite{nam2022filteraugment}, and, for transformer models, frequency warping~\cite{li2024atst-f}. We use the \textit{per-class GRUs} variant (max. learning rate fixed to 1e-3), tune $\lambda_{\mathrm{iou}}$ in $\{0.5, 1.0, 2.0, 4.0\}$, and keep other hyperparameters fixed as specified in Section~\ref{sec:method}. On \textit{AS-Strong-Full}, we follow~\cite{schmid2025effective} but train for 70 epochs, as our method substantially reduces overfitting. Due to the long tail of rare classes, we use the \textit{Single GRU} variant that is trained across all classes. 

\vspace{-5pt}
\subsection{Post-processing}

We compare our method to three baseline approaches: MF, SEBB, and HSM3. For MF and SEBB, hyperparameters are tuned on the AS-Strong-10 validation set after training. For MF, class-wise filter lengths are optimized over a $0\text{--}2\,\mathrm{s}$ grid in 200 ms steps. For SEBB, we use \revaq{change-detection-based SEBB (cSEBB)} and follow the recommended hyperparameter grid from~\cite{ebbers2024sebbs}. For HSM3~\cite{yoshinaga2024onset}, we match their experimental setup and tune the learning rate.

\vspace{-8pt}
\section{Results}

In this section, we present three sets of results. First, we evaluate the overall impact of our method (Section~\ref{subsec:res_as_strong_10}). Next, we analyze the contribution of each individual component (Section~\ref{subsec:config_study}). Finally, we scale to all AS-Strong classes and compare to the state of the art~\cite{schmid2025effective} (Section~\ref{subsec:results_as_strong_full}).

\vspace{-10pt}
\subsection{Results on AS-Strong-10}
\label{subsec:res_as_strong_10}

Table~\ref{tab:main} compares MF and SEBB post-processing for models trained with frame-wise BCE loss, alongside HSM3~\cite{yoshinaga2024onset} and our proposed method (see Section~\ref{sec:method}). SEBB consistently outperforms MF across models and metrics, aligned with the results in~\cite{ebbers2024sebbs}. HSM3 matches SEBB performance without the need for post-processing hyperparameter tuning but adds substantial computational complexity, as reported in~\cite{yoshinaga2024onset}. Our method delivers clear, consistent improvements over SEBB and HSM3 for all models and metrics, especially for the CRNN, which sees a 16\% relative PSDS1 increase. Each class-wise GRU adds only 26K parameters (totaling 260K for 10 classes in \textit{AS-Strong-10}). Notably, the CRNN with our method (with 1.4M parameters) matches transformer models (BEATs, ATST-F; $\approx$90M parameters) using MF post-processing, i.e., a 60x reduction in model size achieved through better temporal modeling, highlighting the critical role of post-processing.

\begin{table}[t!]
\centering
\caption{Performance of our method (\textbf{Ours}) versus \textbf{MF}, \textbf{SEBB}, and \textbf{HSM3}.}
\begin{tabular}{l|cc|cc|cc|cc}
\hline
\textbf{Model} 
& \multicolumn{2}{c|}{\textbf{MF}} 
& \multicolumn{2}{c|}{\textbf{SEBB}} 
& \multicolumn{2}{c|}{\textbf{HSM3}}
& \multicolumn{2}{c}{\textbf{Ours}} \\
\cline{2-9}
& P1 & F1
& P1 & F1
& P1 & F1
& P1 & F1 \\
\hline
CRNN     & 36.9 & 30.4 & 41.1 & 32.3 & 39.8 & 33.2 & \textbf{48.0} & \textbf{40.6} \\
MN-GRU   & 41.4 & 33.9 & 45.7 & 37.2 & 45.1 & 38.8 & \textbf{49.5} & \textbf{42.5} \\
BEATs    & 48.4 & 40.2 & 52.8 & 44.0 & 52.5 & 44.5 & \textbf{55.2} & \textbf{46.7} \\
ATST-F   & 48.2 & 39.9  & 51.9 & 42.4 & 52.3 & 44.9 & \textbf{56.6} & \textbf{48.9} \\
\hline
\end{tabular}
\label{tab:main}
\end{table}

\vspace{-5pt}
\begin{table}[t]
\centering
\caption{Assessment of method components in a configuration study.}
\begin{tabular}{l|l|
>{\columncolor{gray!20}}c
>{\columncolor{gray!20}}c
>{\columncolor{gray!20}}c
>{\columncolor{blue!10}}c}
\hline
\textbf{Model} & \textbf{Metric} 
& \cellcolor{gray!20}BL 
& \cellcolor{gray!20}+RED 
& \cellcolor{gray!20}+OOL 
& \cellcolor{blue!10}+EPN \\
\hline
\multirow{2}{*}{CRNN} 
  & PSDS1 & 41.1 & 42.9 & 46.2 & \textbf{48.0} \\
  & cF1   & 32.3 & 30.4 & 39.7 & \textbf{40.6} \\
\hline
\multirow{2}{*}{ATST-F} 
  & PSDS1 & 51.9 & 52.1 & 53.4 & \textbf{56.6} \\
  & cF1   & 42.4 & 43.8 & 46.0 & \textbf{48.9} \\
\hline
\end{tabular}
\label{tab:config}
\vspace{5pt}
\end{table}

\begin{table}[t]
\centering

\caption{Method comparison on AudioSet Strong classes (PSDS1).}

\begin{tabular}{l|c|c|c}
\hline
\textbf{Method} & \textbf{KD Pipeline} & \textbf{ATST-F} & \textbf{BEATs} \\
\hline
Li et al.~\cite{li2024atst-f}      & \ding{55} & 40.9 & 36.5 \\
Schmid et al.~\cite{schmid2025effective}  & \ding{55} & 41.8 & 44.1 \\
Schmid et al.~\cite{schmid2025effective}  & \ding{51} & 45.8 & 46.5 \\
Ours & \ding{55} & \textbf{47.7} & \textbf{49.6} \\
\hline
\end{tabular}
\label{tab:full}
\vspace{5pt}
\end{table}

\vspace{-5pt}
\subsection{Configuration Study on AS-Strong-10}
\label{subsec:config_study}

We assess the impact of our method's components using the simplest (CRNN) and best-performing (ATST-F) models from Table~\ref{tab:main}. The \textit{BL} column shows models trained with traditional frame-wise BCE loss on presence probabilities. Components are introduced sequentially, aligned with their presentation in Section~\ref{sec:method}. Columns in light gray show SEBB results; the light blue column shows results using Alg.~\ref{alg:region_prop_inference} with the proposed EPNs (\textit{+EPN}).

Table~\ref{tab:config} shows that the main performance gains come from the focal loss on onset/offset probabilities (\textit{+OOL}) and the EPNs with their inference algorithm (\textit{+EPN}). The benefit of each component varies by model: \textit{+OOL} gives the largest boost for CRNN, while ATST-F benefits most from \textit{+EPN} in PSDS1. The RED layer (\textit{+RED}) mainly enables subsequent components with negligible computational overhead.

\vspace{-5pt}
\subsection{Results on AS-Strong-Full}
\label{subsec:results_as_strong_full}

We evaluate our method on all 447 classes of AS-Strong and compare to related work in Table~\ref{tab:full}. Both Li et al.~\cite{li2024atst-f} and Schmid et al.~\cite{schmid2025effective} use MF with a fixed filter length for all classes, due to the lack of a validation set for tuning post-processing hyperparameters. In contrast, our EPNs are optimized end-to-end during training. The \textit{KD Pipeline} column refers to the ensemble knowledge distillation (KD) setup from~\cite{schmid2025effective}, where an ensemble of 15 transformer models is used to boost single-model performance. Table~\ref{tab:full} shows that our method achieves substantial performance gains for both \mbox{ATST-F} and BEATs compared to prior work, even outperforming models trained with the KD pipeline~\cite{schmid2025effective}, thus avoiding the complexity of ensemble distillation. These gains are largely attributable to our additional losses, which act as effective regularization and prevent the overfitting observed in~\cite{schmid2025effective}. The transformer backbones comprise around 90 million parameters, while the \textit{Single GRU} EPNs, using a hidden dimension of 256, add 4.1 million parameters.


\vspace{-5pt}
\section{Conclusion}

In this paper, we present a novel method for temporal event detection applied to SED that enables more accurate temporal localization of events. Our approach is fully compatible with traditional frame-wise models, yet eliminates the need for temporal post-processing and associated hyperparameter tuning. We introduce the RED layer to disentangle onset and offset probabilities, apply losses to onsets and offsets, and propose Event Proposal Networks with a dedicated inference algorithm to directly obtain event regions. Our method yields significant performance gains over related works on subsets of AudioSet Strong and, when scaled to all classes, achieves a new state-of-the-art PSDS1 score of 49.6, surpassing the previous best of 46.5. A current limitation is the lack of real-time inference \rev{capability}, which we aim to address in future work.

\bibliographystyle{IEEEtran}
\bibliography{references}

\end{document}